\documentclass{ws-ijmpe}
\usepackage[super,compress]{cite}
\usepackage{color}
\begin{document}

\markboth{Y. Yang, H. Hassanabadi, H. Chen, Z. W. Long}{DKP oscillator in the presence of a spinning cosmic string}

\catchline{}{}{}{}{}

\title{ DKP oscillator in the presence of a spinning cosmic string}

\author{Yi Yang$^1$\footnote{gs.yangyi17@gzu.edu.cn}, Hassan Hassanabadi$^2$\footnote{h.hasanabadi@shahroodut.ac.ir}, Hao Chen$^1$\footnote{gs.ch19@gzu.edu.cn}, Zheng-Wen Long$^1$\footnote{zwlong@gzu.edu.cn (Corresponding author)}}

\address{$^1$College of Physics, Guizhou University, Guiyang, 550025, China\\
$^2$Faculty of Physics, Shahrood University of Technology, Shahrood, Iran}

\maketitle

\begin{history}
\received{26 March 2021}
\revised{20 May 2021}
\accepted{26 May 2021}
\end{history}

\begin{abstract}
We investigate the modification of gravitational fields generated by topological defects on a generalized Duffin-Kemmer-Petiau (DKP) oscillator for spin-0 particle under spinning cosmic string background. The generalized DKP oscillator equation under spinning cosmic string background is established, and the impact of the Cornell potential on the generalized DKP oscillator is presented. We give the influence of space-time and potential parameters on energy levels.
\end{abstract}


\ccode{PACS numbers: 03.65.Pm, 98.80.Cq, 03.65.Ge.}

\section{Introduction}
In field theory, the DKP equation is usually used to describe spin-zero and spin-one field, which is a linear wave
equation\cite{non22,duffin,non19,kemmer2,castlb}. Its mathematical structure is similar to the Dirac equation, and an obvious difference is that the gamma matrix is replaced by the beta matrix satisfying the DKP algebra. The DKP algebra has three common irreducible representations, namely a trivial one-dimensional representation, a five-dimensional representation describing spin-zero particles and field, and a ten-dimensional representation describing spin-one particles and field\cite{irre1,irre2}. When the DKP theory is used to study the interaction between the meson and nucleus, it can better explain the experimental data than the Klein-Gordon (KG) equation\cite{fried}.

Recently, theoretical physicists are paying more and more attention to DKP oscillator \cite{boum1,boum2,hh1,gomez}. In Ref. \cite{advance8}, the authors have analysed the DKP oscillator under the momentum space. The DKP oscillator for spins-zero and spins-one bosons has been investigated in noncommutative space \cite{falek}. The DKP oscillator in curved space-time attracted a lot of attention. In Ref. \cite{chen1}, the authors have  investigated generalized DKP oscillator under chiral conical background. The DKP oscillator in non-inertial frame of reference has been studied by Castro \cite{castro}. The DKP oscillator under G\"{o}del-type background has been investigated by Ahmed \cite{fa1}. The DKP oscillator under the cosmic string background and spinning cosmic string (SCS) background have been analysed by \cite{hh1} and \cite{base}, respectively. In Ref. \cite{screw}, the authors have examined the DKP oscillator when the  existence of screw dislocation considering two different potential functions. It is worth noting that Furtado et al. are the pioneers of dynamics in the presence of screw dislocations. They discussed many relativistic and non-relativistic dynamics in the presence of screw dislocations \cite{fc1,fc2,fc3,fc4,fc5}.

The study about the generalized oscillator on curved space-time plays an increasingly crucial character. In Ref. \cite{fa2}, the author has studied the generalized KG oscillator by means of the Kaluza-Klein theory. Carvalho et al. studied KG oscillator in the background of a topological defect in a Kaluza-Klein theory \cite{fc6}. The energy spectrum and
wave function of the KG operator are studied considering influence of topological defects in G\"{o}del-type
spacetimes \cite{fc7}. In Ref. \cite{fa3}, the author has studied the generalized KG oscillator under spinning cosmic string background. We have investigated the generalized Dirac oscillator under cosmic dislocation background \cite{chen2}. The Aharonov-Bohm effect on the generalized KG oscillator has been studied \cite{lut}. The quantum dynamics of the generalized DKP oscillator under spinning cosmic strings background has not been established. Meanwhile, DKP oscillator can describe spin-zero and spin-one particles. Therefore, it will be a meaningful work to study the relativistic quantum dynamics of the generalized DKP oscillator in the spinning cosmic strings spacetime.

\section{The generalized DKP oscillator under SCS background} \label{sec:result}

Cosmic strings, as a topologically stable gravitational defects, have attracted widespread attention \cite{36eo,27fa,72da,53kk,28kk,54kk}. In this work we focus on spinning cosmic string. The line element of spinning cosmic string background in cylindrical coordinates can be written as\cite{mazur,gal,bekenstein,bezer,muniz,cunha}
\begin{equation}
\begin{aligned}
d s^{2} &=-(d t+a d \varphi)^{2}+d r^{2}+\alpha^{2} r^{2} d \varphi^{2}+d z^{2} \\
&=-d t^{2}+d r^{2}-2 a d t d \varphi+\left(\alpha^{2} r^{2}-a^{2}\right) d \varphi^{2}+d z^{2},
\end{aligned}
\end{equation}
where $-\infty<z<\infty, r \geq 0,$ and $0 \leq \varphi \leq 2 \pi$, and we use the natural units ($\hbar$= 1, $c$ = 1). The range of angular parameter $\alpha$ is (0, 1]. We make $a= 4GJ$, where $G$ denotes the universal
gravitation constant and $J$ denotes the angular momentum of the spinning string. Thus, the constant parameter $a$ corresponds to the spinning cosmic string.

The DKP equation under SCS background is given by \cite{hh1,hh2}
\begin{equation}\label{eq2}
\left(i \beta^{\mu} \nabla_{\mu}-M\right) \Psi=0.
\end{equation}
In the above formula, the covariant derivative can be written as
\begin{equation}
\nabla_{\mu}=\partial_{\mu}+\Gamma_{\mu}(x),
\end{equation}
where $\Gamma_{\mu}$ is the spinor connection. It can be read as
\begin{equation}\label{eq4}
\Gamma_{\mu}=\frac{1}{2} \omega_{\mu a b}\left[\beta^{a}, \beta^{b}\right].
\end{equation}
The beta-matrices $\beta^{a}$ and $\beta^{b}$ in Minkowski spacetime, are the standard Kemmer matrices, meet the DKP algebra
\begin{equation}
\beta^{a} \beta^{c} \beta^{b}+\beta^{b} \beta^{c} \beta^{a}=\beta^{a} g^{c b}+\beta^{b} g^{c a}.
\end{equation}
The spin affine connection $\omega_{\mu a b}$ of the Eq. (\ref{eq4}) are given by
\begin{equation}
\omega_{\mu a b}=\eta_{a c} e_{v}^{c} e_{b}^{\sigma} \Gamma_{\sigma \mu}^{v}-\eta_{a c} e_{b}^{v} \partial_{\mu} e_{v}^{c}.
\end{equation}
When we consider the situation of curved space-time, DKP matrices $\beta^{\mu}$ of the Eq. (\ref{eq2}) are obtained by
\begin{equation}\label{eq7}
\beta^{\mu}=e_{a}^{\mu} \beta^{a}
\end{equation}
The standard DKP matrices in Minkowski space-time can be expressed as
\begin{equation}
\beta^{0}=\left(\begin{array}{cc}
\Sigma & 0_{2 \times 3} \\
0_{3 \times 2} & 0_{3 \times 3}
\end{array}\right), \quad  \boldsymbol \beta=\left(\begin{array}{cc}
0_{2 \times 2} & \Pi \\
-\Pi^{T} & 0_{3 \times 3}
\end{array}\right),
\end{equation}
where
\begin{equation}
\begin{array}{cc}
\Sigma=\left(\begin{array}{ll}
0 & 1 \\
1 & 0
\end{array}\right), & \Pi^{1}=\left(\begin{array}{rrr}
-1 & 0 & 0 \\
0 & 0 & 0
\end{array}\right), \\[10pt]
\Pi^{2}=\left(\begin{array}{rrr}
0 & -1 & 0 \\
0 & 0 & 0
\end{array}\right), & \Pi^{3}=\left(\begin{array}{rrr}
0 & 0 & -1 \\
0 & 0 & 0
\end{array}\right).
\end{array}
\end{equation}
For the line element we consider, we can choose the following tetrads
\begin{equation}
\begin{aligned}
e_{a}^{\mu}=\left(\begin{array}{cccc}
1 & \frac{a \sin \varphi}{r \alpha} & \frac{-a \cos \varphi}{r \alpha} & 0 \\
0 & \cos \varphi & \sin \varphi & 0 \\
0 & \frac{-\sin \varphi}{r \alpha} & \frac{\cos \varphi}{r \alpha} & 0 \\
0 & 0 & 0 & 1
\end{array}\right).
\end{aligned}
\end{equation}
According to the above specific tetrad and Eq. (\ref{eq7}),
the beta-matrices in the curved-space can be derived
\begin{equation}
\begin{aligned}
&\beta^{t}=e_{a}^{t} \beta^{a}=\beta^{0}-\frac{a}{r \alpha} (\cos \varphi \beta^{2}-\sin \varphi \beta^{1}), \\
&\beta^{r}=e_{a}^{r} \beta^{a}=\cos \varphi \beta^{1}+\sin \varphi \beta^{2}, \\
&\beta^{\varphi}=e_{a}^{\varphi} \beta^{a}=\frac{1}{r \alpha}(\cos \varphi \beta^{2}-\sin \varphi \beta^{1}), \\
&\beta^{z}=e_{a}^{z} \beta^{a}=\beta^{3},
\end{aligned}
\end{equation}
and the spin connections can be given by
\begin{equation}
\Gamma_{\varphi}=(1-\alpha)\left[\beta^{1}, \beta^{2}\right].
\end{equation}
The DKP oscillator could be derived through the substitution
\begin{equation}\label{eq13}
i \nabla \rightarrow i \nabla-i M \omega \eta^{0} \mathbf{r},
\end{equation}
where $\omega$ represents the frequency of oscillator , $\eta^{0}=2\left(\beta^{0}\right)^{2}-1$ and $\mathbf{r} = (0, r, 0, 0)$. In this paper, we focus on the generalized oscillator under the spinning cosmic string background. The generalized oscillator is introduced by the substitution that $r$ in Eq. (\ref{eq13}) is substituted by $f(r)$, i.e. the vector $\mathbf{r}$ transforms into $\mathbf{r} = (0, f(r), 0, 0)$.
Therefore, the equation of the generalized DKP oscillator in curved space-time can be derived
\begin{equation}\label{eq14}
\left[i \beta^{\mu} \partial_{\mu}+i \beta^{\mu} \Gamma_{\mu}+i \beta^{r} M \omega \eta^{0} f(r)-M\right] \Psi(x)=0.
\end{equation}

We know that the interaction is time-independent. Therefore, the DKP spinor could be read as
\begin{equation}\label{eq15}
\displaystyle\Psi(t, r, \varphi, z)=e^{-i E t+i m \varphi+i k z}\left(\begin{array}{c}\Phi_{1}(r) \\ \Phi_{2}(r) \\ \Phi_{3}(r) \\ \Phi_{4}(r) \\ \Phi_{5}(r)\end{array}\right),
\end{equation}
with $E$ being the energy, $m$ being the magnetic quantum number, and $k$ being the wave number.
Substituting Eq. (\ref{eq15}) into Eq. (\ref{eq14}), we can derive the following five algebraic equations
\begin{equation}
\begin{aligned}
&r \alpha\left(-M \Phi_{1}(r)+E \Phi_{2}(r)+k \Phi_{5}(r)+\cos \varphi(a E+m) \Phi_{4}(r)\right. \\
&+\left(i \cos \varphi\left(\Phi_{3}(r)+\alpha(-1+r M \omega f(r)) \Phi_{3}(r)-r \alpha \Phi_{3}^{\prime}(r)\right)\right) \\
&\left.-\sin \varphi\left((a E+m) \Phi_{3}(r)-i(1-\alpha+r \alpha M \omega f(r)) \Phi_{4}(r)\right.\right.\\
&\left.\left.+i r \alpha \Phi_{4}^{\prime}(r)\right)\right)=0,\\[7pt]
&E \Phi_{1}(r)-M \Phi_{2}(r)=0,\\[7pt]
&(a E+m) \sin \varphi \Phi_{1}(r)-M r \alpha \Phi_{3}(r)\\
&+ir\alpha \cos \varphi\left(M\omega f(r) \Phi_{1}(r)+\Phi_{1}^{\prime}(r)\right)=0,\\[7pt]
&(a E+m) \cos \varphi \Phi_{1}(r)+Mr \alpha \Phi_{4}(r)\\
&-ir \alpha \sin \varphi\left(M\omega f(r) \Phi_{1}(r)+\Phi_{1}^{\prime}(r)\right)=0,\\[7pt]
&k \Phi_{1}(r)+M \Phi_{4}(r)=0.
\end{aligned}
\end{equation}
Therefore, the generalized DKP oscillator in the spinning cosmic string background can be read as
\begin{equation}\label{secondorder}
\begin{aligned}
&\frac{\mathrm{d}^2}{\mathrm{d}r^2}\Phi_{1}(r)+\frac{\alpha-1}{r\alpha} \frac{\mathrm{d}}{\mathrm{d}r}\Phi_{1}(r)+\left[E^2-M^{2}-k^{2}\right.\\
&\left.-M^2 \omega^2f^2(r)+\frac{M\omega f(r)}{r}
-\frac{M \omega f(r)}{r\alpha}-\frac{(a E+m)^{2}}{r^{2} \alpha^{2}}+M \omega \frac{\mathrm{d}}{\mathrm{d}r}f(r)\right] \Phi_{1}(r)=0.
\end{aligned}
\end{equation}
\section{The generalized DKP oscillator by considering the Cornell Potential} \label{sec:result}
In this section, we study the effect of the Cornell potential on the generalized DKP oscillator. The Cornell potential was used to describe the interaction between heavy quarks, which is the sum of Coulomb potential and linear potential \cite{fc8}. It can be written as \cite{72da,cornel3,cornel1,rll,ijmpa,mpla}
\begin{equation}
f(r)=\Delta _1 r+\frac{\Delta _2}{r},
\end{equation}
where $\Delta _1$ and $\Delta _2$ is potential parameter. Substituating this potential into Eq. (\ref{secondorder}) we can get
\begin{equation}
\begin{aligned}
&\frac{\mathrm{d}^{2}}{\mathrm{~d} r^{2}} \Phi_{1}(r)+\frac{\alpha-1}{r \alpha} \frac{\mathrm{d}}{\mathrm{d} r} \Phi_{1}(r)+\left[E^{2}-M^{2}-k^{2}\right.\\
&\left.-\Delta_{1}^{2} M^{2} \omega^{2} r^{2}-\frac{\Delta_{2} M \omega}{r^{2} \alpha}-\frac{\Delta_{1} M \omega}{\alpha}-\frac{(a E+m)^{2}}{r^{2} \alpha^{2}} \right.\\
&\left.+2 \Delta_{1} M \omega-2 \Delta_{1} \Delta_{2} M^{2} \omega^{2}-\frac{\Delta_{2}^{2} M^{2} \omega^{2}}{r^{2}}\right] \Phi_{1}(r)=0.
\end{aligned}
\end{equation}
Then, we assume
\begin{equation}
\begin{aligned}
\Phi_{1}=r^{\frac{1}{2\alpha }}\zeta_{n, \ell}(r).
\end{aligned}
\end{equation}
Thus, the Eq.(19) becomes
\begin{equation}
\begin{aligned}
&\frac{\mathrm{d}^2\zeta_{n, m}(r)}{\mathrm{d}r^2}+\frac{1}{r} \frac{\mathrm{d}\zeta_{n, m}(r)}{\mathrm{d}r}+\left[E^2-M^{2}-k^{2}\right.\\&\left.-\Delta _1^2M^2 \omega^2r^2
-\frac{\Delta _2M\omega}{r^2 \alpha}-\frac{1}{4r^2 \alpha^2}-\frac{\Delta _1M \omega}{\alpha}\right.\\&\left.-\frac{(a E+m)^{2}}{r^{2} \alpha^{2}}+2\Delta _1M\omega
-2\Delta _1\Delta _2M^2\omega^2\right.\\&\left.-\frac{\Delta _2^2M^2\omega ^2}{r^2} \right] \zeta_{n, m}(r)=0.
\end{aligned}
\end{equation}

When we change the variable as $\rho = r^2$, Eq. (21) can be written as
\begin{equation}
\begin{aligned}
&\frac{\mathrm{d}^2\zeta_{n, m}(\rho)}{\mathrm{d}\rho^2}+\frac{1}{\rho} \frac{\mathrm{d}\zeta_{n, m}(\rho)}{\mathrm{d}\rho}+\frac{1}{4\rho^2}\left[(E^2-M^{2}-k^{2}\right.\\
&\left.-\frac{\Delta _1M \omega}{\alpha}+2\Delta _1M\omega-2\Delta _1\Delta _2M^2\omega^2)\rho\right.\\
&\left.-\Delta _1^2M^2 \omega^2\rho^2 -(\Delta _2^2M^2\omega^2+\frac{1}{4\alpha ^2}+\frac{(aE+m)^2}{\alpha ^2}\right.\\
&\left.+\frac{\Delta _2M\omega}{\alpha}) \right] \zeta_{n, m}(\rho)=0.
\end{aligned}
\end{equation}
By analyzing Eq. (22), we can find that it can be solved by the Nikiforov-Uvarov (NU) method \cite{NUmethod,sedagh,naderi,zhang,qiang,ikot,Hassanabadi,cpb2020}.
Therefore, we can gain these parameters
\begin{equation}\begin{aligned}
\Xi_{1} &=\frac{1}{4}M^{2} \omega^{2}\Delta _1^2, \\
\Xi_{2} &=\frac{E^2\!-M^{2}\!-k^{2}\!-\frac{\Delta _1M \omega}{\alpha}\!+2\Delta _1M\omega \!-2\Delta _1\Delta _2M^2\omega^2}{4} \\
\Xi_{3} &=\frac{1}{4}(\Delta _2^2M^2\omega^2+\frac{1}{4\alpha ^2}+\frac{(aE+m)^2}{\alpha ^2}+\frac{\Delta _2M\omega}{\alpha}), \\
c_{1} &=1, \ c_{2} =c_{3}= c_{4} = c_{5}=0,
 c_{6} =\Xi_{1},
 c_{7} =-\Xi_{2},\\
\quad c_{8} &=\Xi_{3},
\qquad \ \ \, c_{9} =\Xi_{1},
\quad \quad \ \, c_{10} =1+2\sqrt{\Xi_{3}},\\
c_{11} &=2\sqrt{\Xi_{1}},
\quad c_{12} =\sqrt{\Xi_{3}},
\quad \ \, c_{13} =-\sqrt{\Xi_{1}}.
\end{aligned}
\end{equation}

\begin{figure}[b!]
\vspace{0.1cm}
\begin{center}
\includegraphics[width=3.4in,height=2.5in]{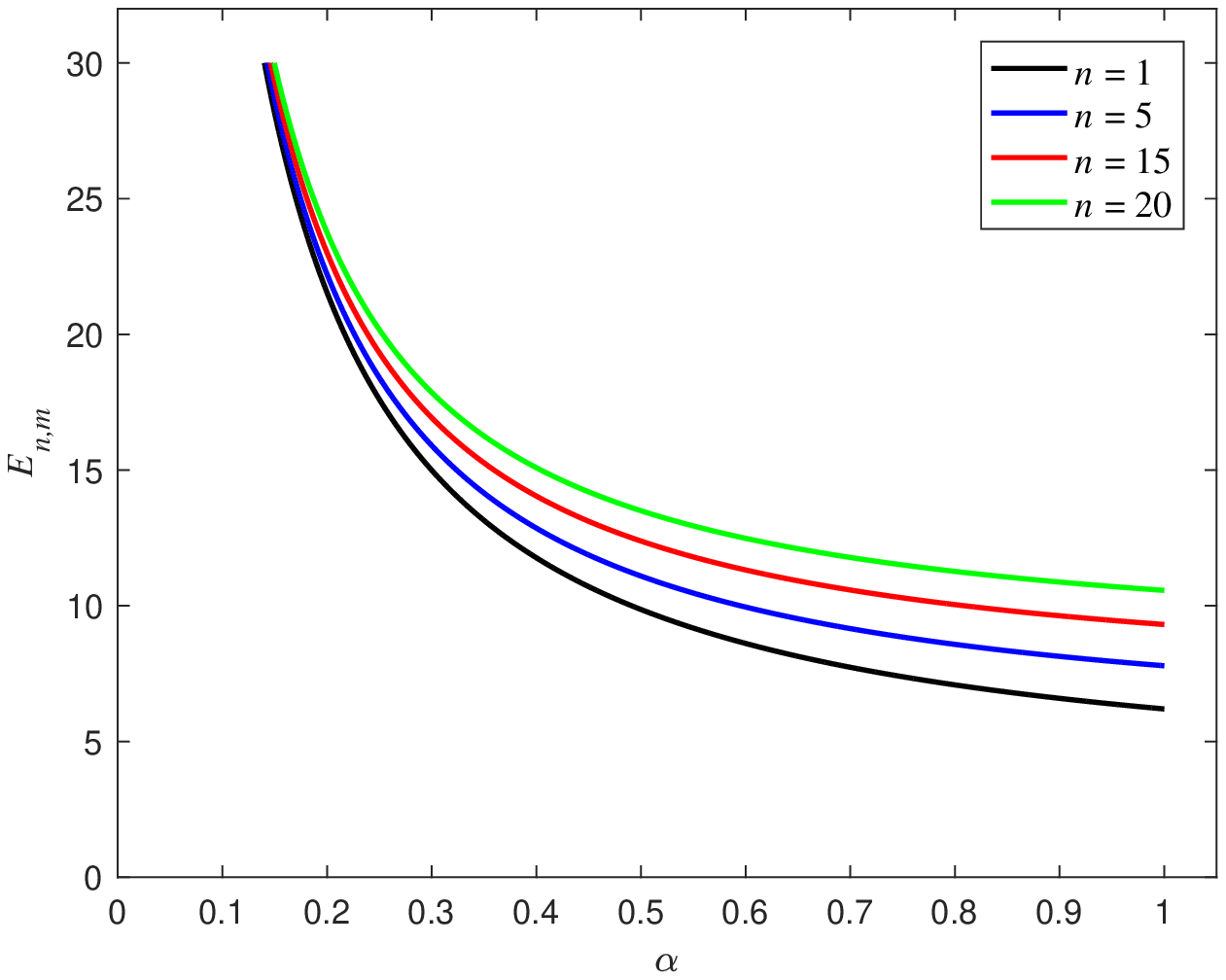}
\end{center}
\setlength{\abovecaptionskip}{-0.3cm}
\caption{Energy levels as a function of $\alpha$ of generalized DKP oscillator for spin-0 particle under the SCS background with  $a=m=k=M=\omega=\Delta_1=\Delta_2=1$.}
\label{Ealpha}
\vspace{-0.2cm}
\end{figure}

\begin{figure}[t!]
\vspace{-0.2cm}
\begin{center}
\includegraphics[width=3.4in,height=2.5in]{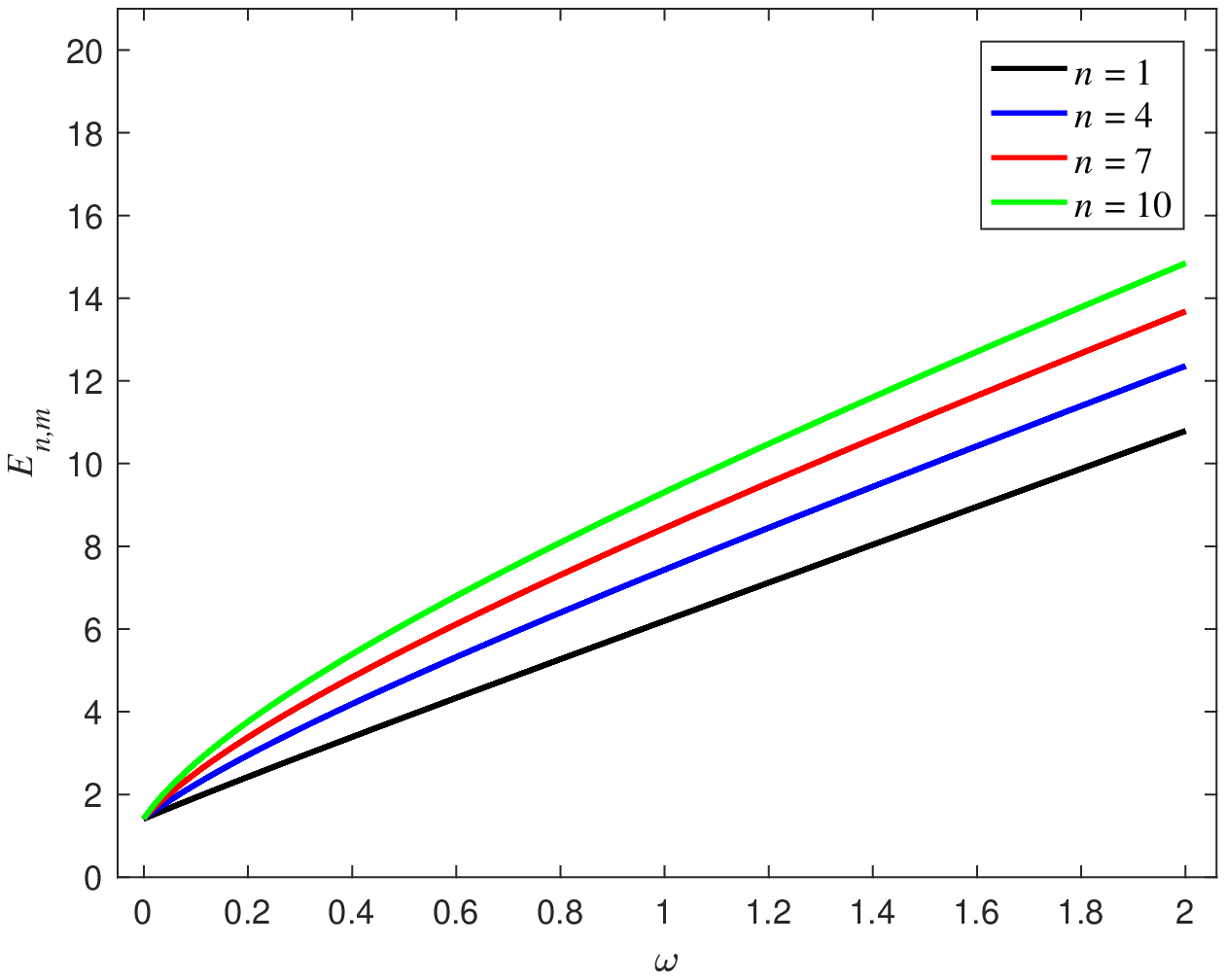}
\end{center}
\setlength{\abovecaptionskip}{-0.3cm}
\setlength{\belowcaptionskip}{0.5cm}
\caption{Energy levels as a function of $\omega$ of generalized DKP oscillator for spin-0 particle under the SCS background with  $a=m=k=M=\alpha=\Delta_1=\Delta_2=1$.}
\label{Eomega}
\vspace{0.2cm}
\end{figure}

\begin{figure}[t!]
\vspace{-0.2cm}
\begin{center}
\includegraphics[width=3.4in,height=2.5in]{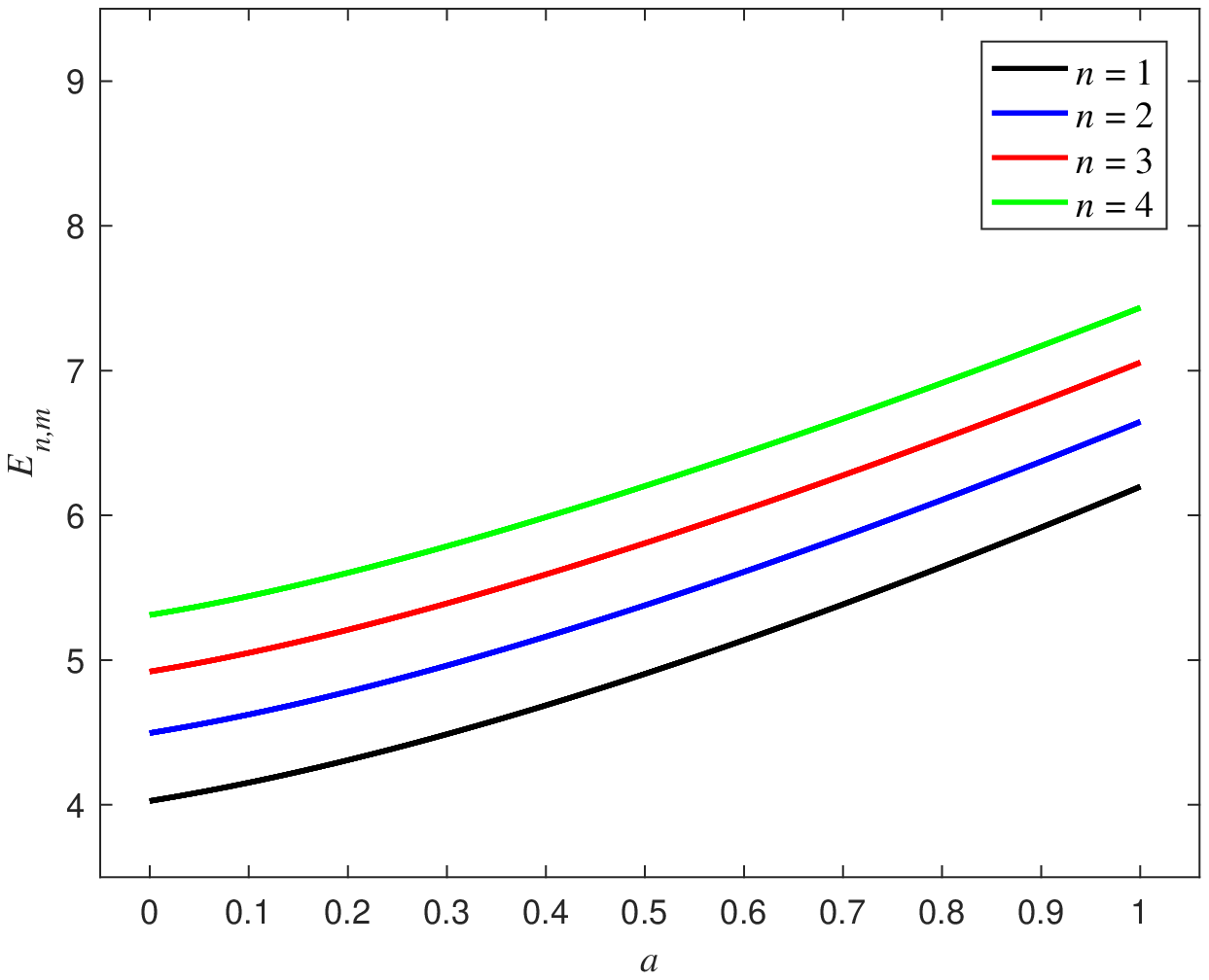}
\end{center}
\setlength{\abovecaptionskip}{-0.3cm}
\caption{Energy levels as a function of $a$ of generalized DKP oscillator for spin-0 particle under the SCS background with $\omega=m=k=M=\alpha=\Delta_1=\Delta_2=1$.}
\label{Ea}
\vspace{0.2cm}
\end{figure}

\begin{figure}[t!]
\vspace{-0.2cm}
\begin{center}
\hspace{-0.7cm}
\includegraphics[scale=0.46]{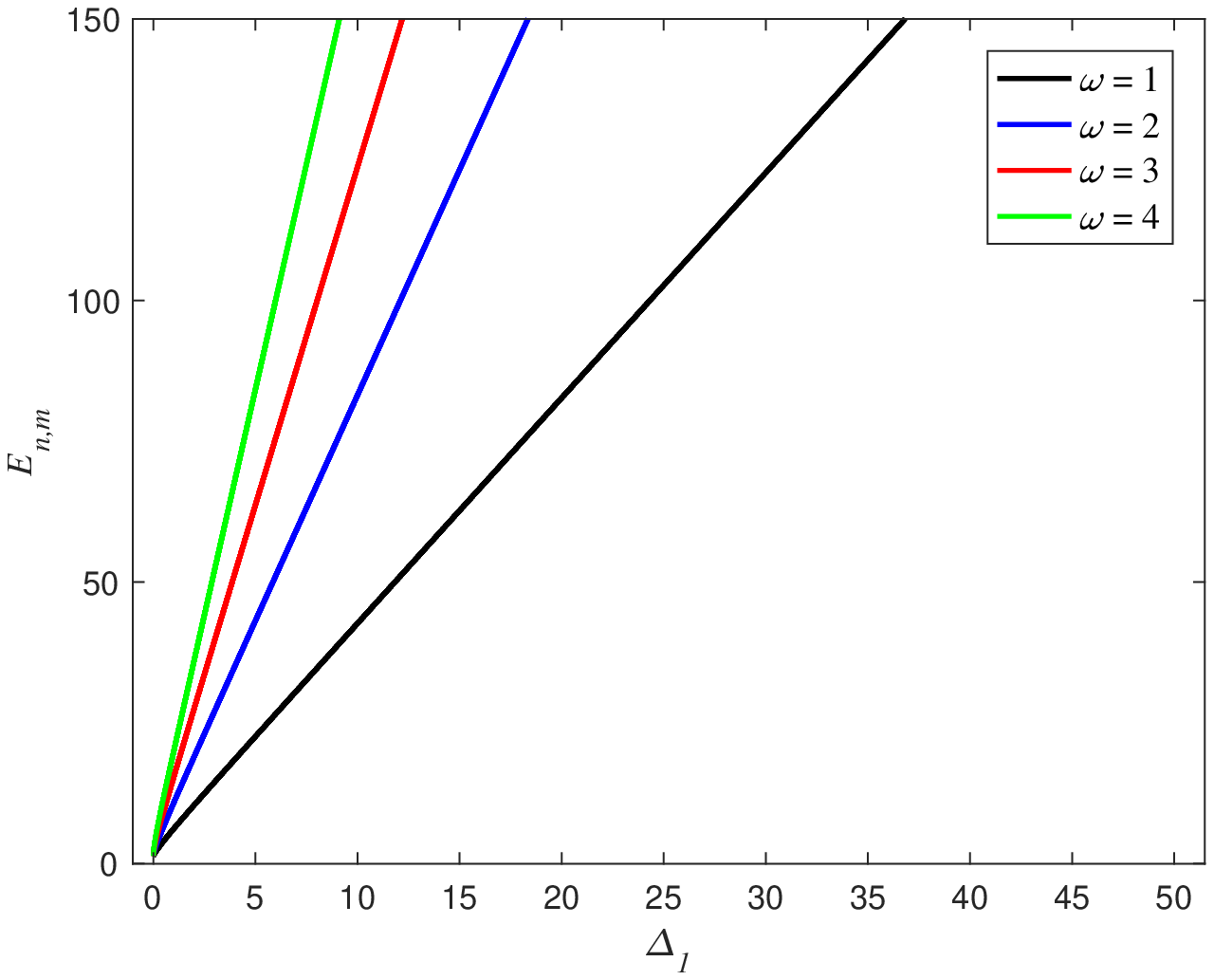}
\hspace{-0.8cm}
\includegraphics[scale=0.46]{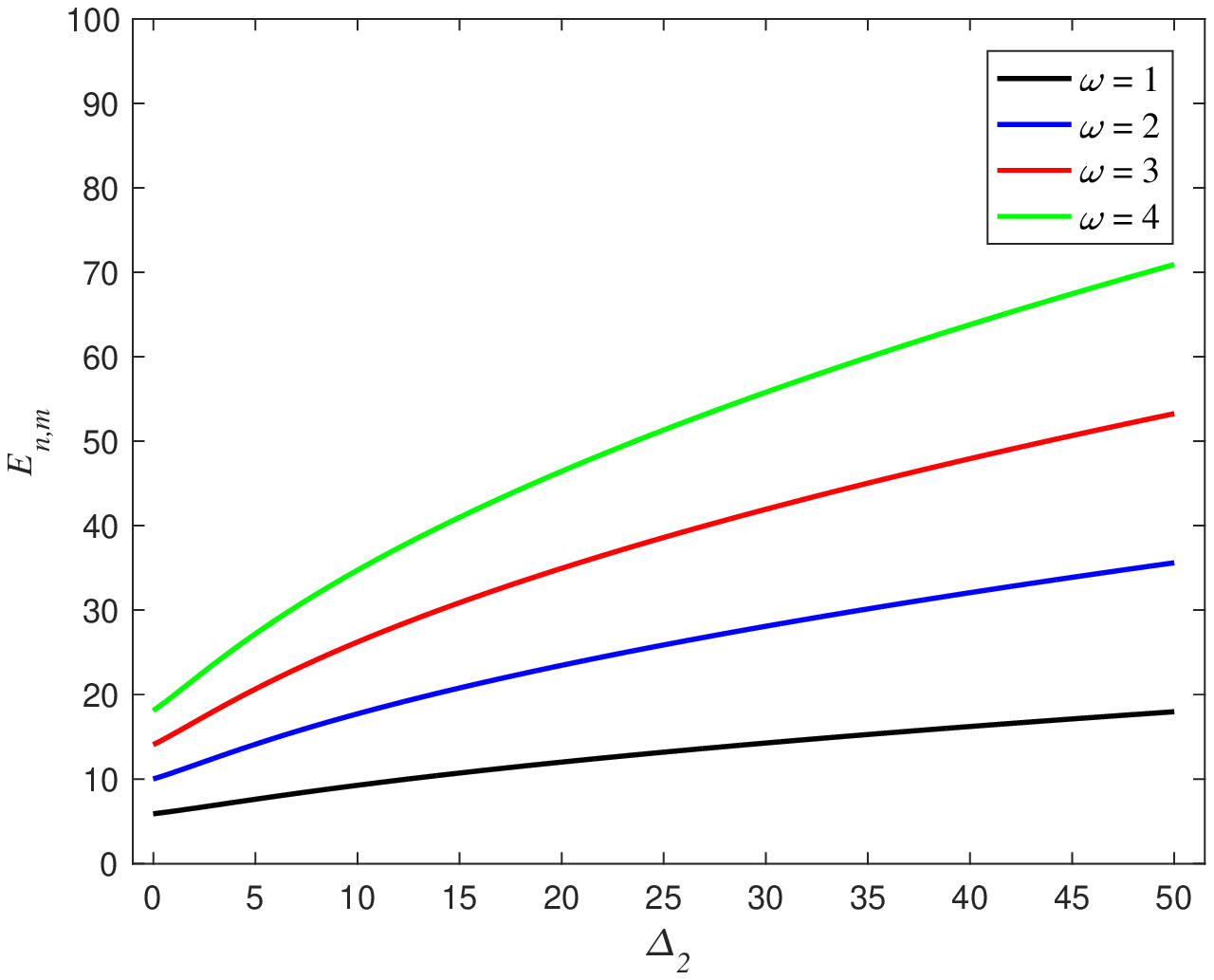}
\end{center}
\setlength{\abovecaptionskip}{-0.3cm}
\setlength{\belowcaptionskip}{0.5cm}
\vspace{0.1cm}
\caption{Energy levels as a function of $\Delta_1$ (with $a=n=m=k=M=\alpha=\Delta_2=1$) and $\Delta_2$ (with $a=n=m=k=M=\alpha=\Delta_1=1$) of generalized DKP oscillator for spin-0 particle under the SCS background.}
\label{Edeta2}
\vspace{0.3cm}
\end{figure}

\begin{figure}[t!]
\vspace{-0.2cm}
\begin{center}
\hspace{-0.7cm}
\includegraphics[scale=0.46]{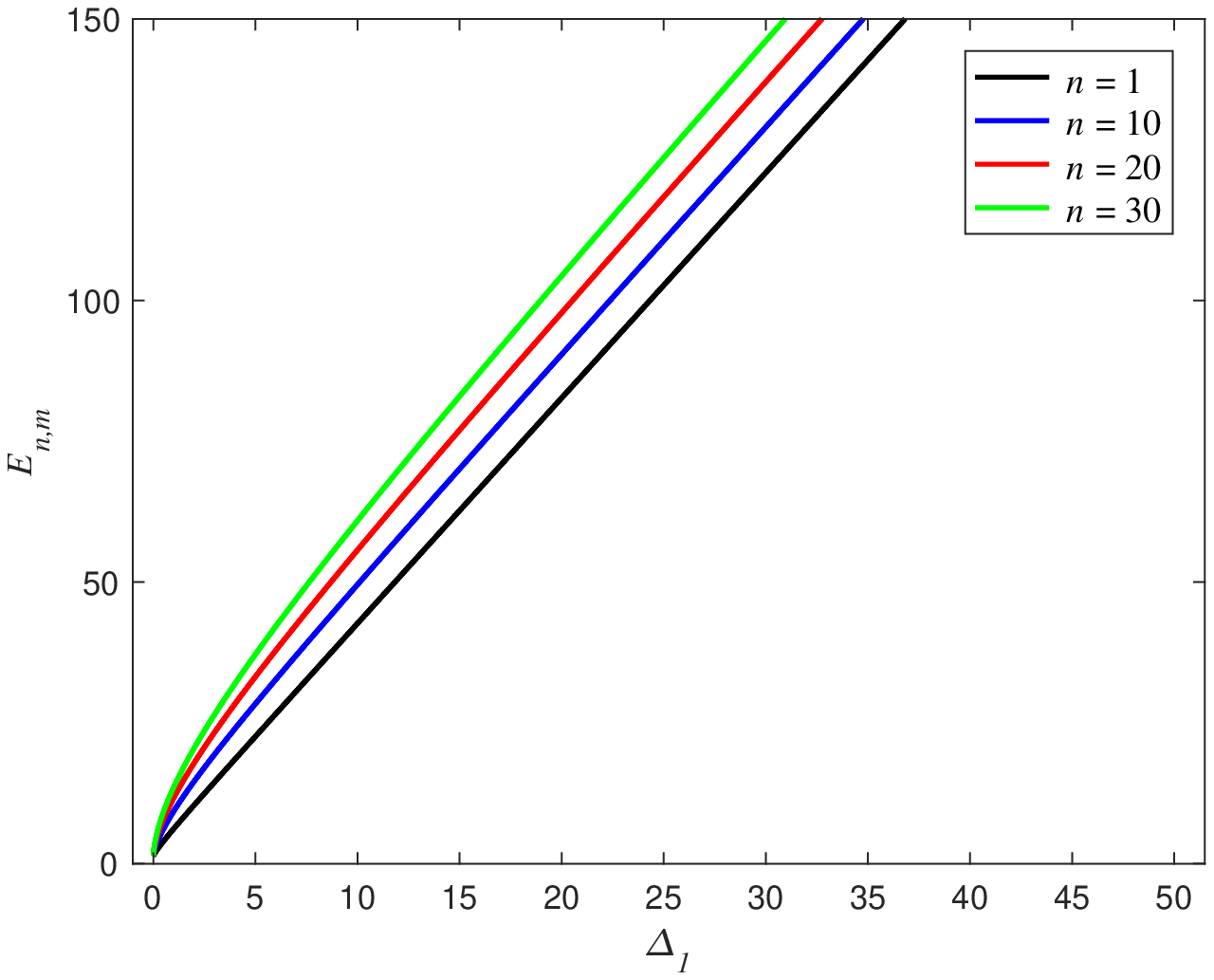}
\hspace{-0.8cm}
\includegraphics[scale=0.46]{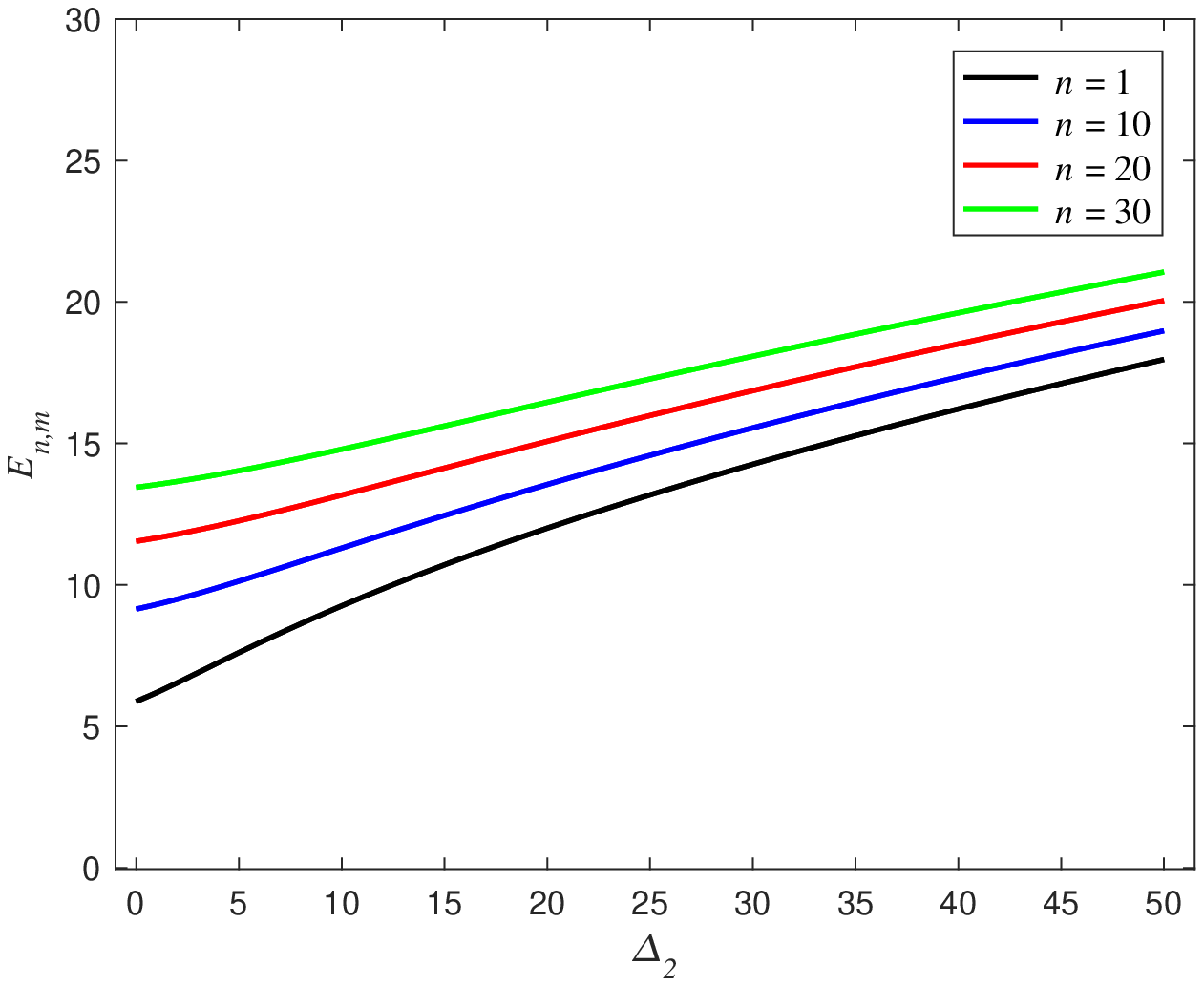}
\end{center}
\setlength{\abovecaptionskip}{-0.3cm}
\setlength{\belowcaptionskip}{0.5cm}
\vspace{0.1cm}
\caption{Energy levels as a function of $\Delta_1$ (with $a=\omega=m=k=M=\alpha=\Delta_2=1$) and $\Delta_2$ (with $a=\omega=m=k=M=\alpha=\Delta_1=1$) of generalized DKP oscillator for spin-0 particle under the SCS background.}
\label{vn}
\vspace{0.3cm}
\end{figure}

On the other hand, the energy spectrum and wave function can be written as
\begin{equation}
\begin{aligned}
&2(2 n+1)\Delta _1M\omega-(E^2-k^2-M^2\\
&+2\Delta _1M\omega-2\Delta _1\Delta _2M^2\omega ^2-\frac{\Delta _1M\omega}{\alpha})+\\
&4\Delta _1M\omega \sqrt{\Delta _2^2M^2\omega^2+\frac{1}{4\alpha ^2}+\frac{(aE+m)^2}{\alpha ^2}+\frac{\Delta _2M\omega}{\alpha}}=0,
\end{aligned}
\end{equation}
and
\begin{equation}
\Phi_{1}(r)=r^{\Theta} \exp \left(-\frac{\Delta _1M\omega r^2}{2}\right) L_{n}^{\Theta}\left(\Delta _1M\omega r^2\right),
\end{equation}
where $\Theta=\Delta _2^2M^2\omega^2+\frac{1}{4\alpha ^2}+\frac{(aE+m)^2}{\alpha^2}+\frac{\Delta _1M\omega}{\alpha}$, and $L_n$ is the Laguerre polynomial.

In order to more intuitively reflect the influence of the various parameters of the system we considered on the energy spectrum, we have drawn the energy spectrum diagram. In Fig. \ref{Ealpha}, we show the energy eigenfunctions as function of $\alpha$ for $a=m=k=M=\omega=\Delta_1=\Delta_2=1$, and it show the energy levels decreases with the increase of $\alpha$. Especially in the small $\alpha$ region, the decreasing trend of the energy spectrum is particularly obvious, while the decrease in the large $\alpha$ region is slower. In Fig. \ref{Eomega}, we show the energy eigenfunctions as function of $\omega$ for $a=m=k=M=\alpha=\Delta_1=\Delta_2=1$. We can obtain the energy spectrum increases as $\omega$. In Fig. \ref{Ea}, we show the energy eigenfunctions as function of $a$ for $\omega=m=k=M=\alpha=\Delta_1=\Delta_2=1$. We can see the energy spectrum increases as $a$. In addition, the trend of increasing in the large $a$ region is more obvious.  In Fig. \ref{Edeta2}, we show the energy eigenfunctions as function of $\Delta_1$ and $\Delta_2$ for $a=n=m=k=M=\alpha=\Delta_2=1$ and $a=n=m=k=M=\alpha=\Delta_1=1$, respectively. We can learn the energy spectrum increases with the increase of $\Delta_1$ and $\Delta_2$, and the energy spectrum is more sensitive to the change of $\Delta_1$. At the same time, we also give the situation of different $n$ in Fig. 5, we can see the physical mechanism similar to Fig. 4. We can conclude that the energy spectrum of the generalized DKP oscillator is strongly dependent on the linear part of the Cornell potential.

\section{Conclusion} \label{sec:result}
In this study, the generalized DKP oscillator subjected to the Cornell potential towards spin-0 particle under the SCS background is investigated. We derive the equation of the generalized DKP oscillator in the spinning cosmic string background, and analyse the effect of the Cornell potential on the generalized DKP oscillator equation in this space-time. By means of the NU method, we have given the energy spectrum and wave function. It is worth noting that when the parameter satisfies $a=0, \Delta_1=1, \Delta_2=0$, the corresponding energy spectrum and wave function are consistent with the results of Ref. \cite{hm}. We also incorporate the influences of the topological defect, spinning cosmic string, and Cornell potential to modify the energy levels and wave function of generalized DKP oscillator for the spin-zero particle through the existence of parameters $\alpha$, $a$, $\Delta_1$ and $\Delta_2$.
 Importantly, the potential function parameter $\Delta_2$ has a non-negligible contribution to our studying system.
Besides, focusing on the Fig. 1, we see that as the $\alpha$ increases from zero to one, the distance between the energy levels increases; meanwhile, in low-value alpha, we go to a degeneracy. The same analysis can be expressed for the $\omega$ changes associated with Fig. 2. According to Fig. 3, by increasing the effect of the spinning cosmic string, or in other words, by increasing the value of the parameter $a$ from zero to one, it can be seen that the values of energy associated with different levels increase at almost the same rate. It should be mentioned that, by increasing the values of the topological defect parameter $\alpha$ and the spinning cosmic string $a$, it can be seen that energy eigenvalues decrease and increase, respectively. In this regard, in Figs. 4 and 5, it can be seen that with increasing the values of the potential parameters, the energy generally increases.

\section*{Acknowledgements}
The authors thank the referee for a thorough reading of our manuscript and for constructive suggestion.
This research was funded by the National Natural Science Foundation of China (Grant No.11465006 and 11565009).

\end{document}